# Effect of Protein Environment on the Shape Resonances of RNA Pyrimidine Nucleobases: Insights From a Model System

Sneha Arora[a], Jishnu Narayanan S J[a], Idan Haritan[b], Amitava Adhikary[c], and Achintya Kumar Dutta[a,d]*

[a] *Department of Chemistry, Indian Institute of Technology Bombay, Powai, Mumbai 400076, India.*

[b] *The Alexander Kofkin Faculty of Engineering, Bar-Ilan University, Ramat Gan, 5290002, Israel.*

[c] *Free Radical and Radiation Biology Program, Department of Radiation Oncology, Carver College of Medicine, University of Iowa, Iowa 52242, United States.*

[d] *Department of Inorganic Chemistry, Faculty of Natural Sciences, Comenius University, Bratislava Ilkovičova 6, Mlynská dolina 842 15 Bratislava, Slovakia*

**Abstract**

In this work, the effect of amino acid environment on the nucleobase-centered anion radical shape resonances is investigated by employing uracil as a model system for pyrimidine base in RNA. Anionic uracil-glycine complexes have been used to model the RNA-protein interactions. The resonance positions and widths of these complexes have been simulated using the equation of motion coupled cluster method coupled with resonance via Padé approach. Our work shows that in the transient negative ion (TNI, or, the anion radical of glycine:uracil complex), glycine stabilizes the nucleobase-centered resonances through hydrogen bonding, increasing the lifetime of TNI. At the same time, a glycine-centered resonance shows the ability of amino acids to capture the electron density and move it away from the uracil nucleobase. At the microsolvation level, this modeling indicates that amino acids would have more influence on nucleobase-centered resonances in the TNI than that displayed by the corresponding aqueous environment.

*achintya@chem.iitb.ac.in

# 1. Introduction

In recent years, there has been a growing interest in understanding collisions between electrons and biomolecules due to their importance in radiation damage and biological applications.[1] However, a complete picture of these processes at the molecular level is still not clear. A detailed understanding of electron–biomolecule collisions is essential for connecting molecular-level interactions to cellular processes. This is particularly significant in biological environments, where ionizing radiation produces a large number of electrons, ions, free radicals, and excited species.[2,3] In the condensed phase, secondary electrons are the key contributors to radiation-induced cellular damage, such as mutations, genomic instability, and apoptosis.[4–8] Radiation-induced DNA and RNA damage includes strand breaks, base modifications, cross-links, and tandem lesions including clustered lesions.[9] Such damage arises either from radiation-mediated direct interactions with DNA/RNA components or through the reactive species generated from surrounding molecules (e.g., water, salts, proteins, oxygen), i.e., through the indirect effect of radiation.

Secondary electrons generated by ionizing radiation rapidly undergo successive inelastic collisions with surrounding molecules, producing a spectrum of medium- and low-energy electrons (LEEs), with LEEs typically ranging from 0 to 20 eV.[8,10–16] At energies below the ionization threshold of a molecular target, LEEs can participate in a variety of interactions with the surrounding genetic material. These include elastic scattering, rotational, vibrational, or electronic excitations, and electron attachment. While the excitations can lead to neutral dissociation processes, the electron attachment may result in dissociative electron attachment (DEA).[17–20] Although, both processes occur at specific resonance energies, corresponding to the energy levels of transient states. DEA is initiated by transient negative ions (TNIs) or resonances, formed upon LEE attachment to the genetic materials.[21–30] These resonance states exist only for a very small, albeit finite duration, typically from a few to several hundred femtoseconds, and are primarily localized on nucleobases or phosphate groups.[31] Woldhuis et al.[32] and Sanche and co-workers[17,22,23,33–41] demonstrated that LEEs can induce DNA single- and double-strand breaks in the condensed phase, even at energies as low as ~0 to 4 eV.[24,42–47] LEE attachment to genetic materials can lead to the formation of either shape or core-excited resonances, which depend on the energy of the incoming electron and the properties of the molecules.[28,29,48–53] Shape resonances usually occur when incident electrons with energies between 0 to 4 eV become temporarily confined by the attractive potential barrier arising due to the interaction between the electron and the neutral molecule.[54] The temporarily bound electron may subsequently undergo decay by tunneling through the potential barrier, which leads to autoionization, or alternatively, it can transfer the excess energy to one of the vibrational modes of the adjacent bonds, which may lead to dissociation. These resonances are classified as either σ* or π* based on the specific molecular orbital involved.[31,43,45,51,54–57] Contrary to shape resonances, the core-excited resonances lie above 3−4 eV, and involve the capture of the incoming electron by an electronically excited state of the molecule, which has a positive electron affinity.[28]

Studying resonances provides valuable insights into the complex many-body effects that govern molecular physics. Yet, their theoretical treatment is challenging due to their non-bound, decaying nature. Because of this metastability, resonances cannot be represented by a single discrete eigenfunction of the physical, Hermitian Hamiltonian,[58–67] and require simultaneous treatment of electron correlation and continuum effects. Hence, conventional methods used for bound-state electronic structure calculations are unsuitable for the calculation of resonances. However, these states can be described as a single eigenstate using non-Hermitian quantum mechanics. The non-Hermitian formulation of quantum mechanics leads to a complex eigenvalue, $E_R - i\Gamma/2$, where the real part $E_R$ denotes the energy or position of the resonance state, and the complex part $\Gamma$ represents the resonance state width, which is inversely proportional to its lifetime.[58]

Despite these difficulties, in recent years there has been a development of ab initio quantum chemical $L^2$ methods for resonances. These methods have significantly advanced the study of resonances in many-electron systems, and include complex scaling (CS),[59–63] complex absorbing potential (CAP),[63–71] and stabilization techniques.[72–75] The benefit of adapting quantum chemical $L^2$ techniques for treating resonances, is that they allow for the use of all already available computational tools in quantum chemistry. Among these methods, the stabilization approach, which applies conventional quantum mechanical methods without any modification[72–75], stands out as one of the straightforward methods for simulating resonances.

The proximity of biomolecules, particularly proteins, to genetic materials (e.g., DNA) significantly influences radiation-induced electron attachment pathways and the extent of consequent damage.[76–78] Solomun and Skalicky[79] demonstrated that single-strand binding proteins can reduce LEE-induced single-strand breaks (ca. 4 times) in a pyrimidine polynucleotide, e.g., in (dT)$_{25}$- that was vacuum irradiated with electrons at 3 eV. They attributed this protection to physical shielding as well as to preferential electron attachment to the protein rather than DNA. Ptasińska and coworkers[80] investigated the impact of ultra-high vacuum electron irradiation (1 eV) on thin films of tetramer GCAT in the presence of glycine and arginine. At low amino acid: GCAT ratios, increased DNA fragmentation and single-strand breaks were observed, likely due to H˙ radicals formed near GCAT. In contrast, at higher amino acid: GCAT ratios, glycine, or arginine molecules form dimers instead of complexes with the tetramer, reducing local H˙ radical formation and thereby minimizing DNA damage. The genetic material present in eukaryotic cells can be both RNA and DNA. There is a growing body of evidence correlating RNA damage to disease, e.g., in Alzheimer's disease, high levels of 8-oxo-guanine RNA levels in cytoplasm is reported.[81,82] It has been shown that RNA is more susceptible to radiation damage than DNA in the cellular environment.[83]

These alterations in RNA could reshape our understanding of gene expression and potentially disease mechanisms.[84] Moreover, owing to the presence of 2′-OH group in the sugar moiety, the consequent vulnerability of RNA to radiation-induced damage can differ between coding and noncoding types, among various noncoding RNA classes, and across different mRNA species.[83] One of the proposed explanations is the extent of RNA binding with proteins, which may offer a protective effect against such damage. Also, the investigations of RNA damage provide important information on its structure and folding kinetics.[85] The uracil nucleobase is the simplest model system possible for pyrimidine RNA-bases. Theoretical studies on bound uracil–amino acid anion radical complexes[86,87] suggest that barrierless proton transfer from the carboxylic group of the amino acid to the nucleobase anion radical forms a neutral hydrogenated radical (UH˙) and a stabilized carboxylate anion ($^-$OOC−R− NH$_2$). This spin–charge separation stabilizes the excess electron and prevents LEE-induced nucleobase fragmentation, suggesting a chemical protective role of amino acids. QM/MM simulations have shown that similar chemical protection is also present in the condensed phase.[52] However, the effect of the amino acid environment on the temporarily bound anionic states is not well-understood.

A widely accepted mechanism for the formation of electron-mediated RNA strand break involves the initial attachment of LEEs to nucleobases, forming π* shape resonances,[5,45,48] followed by electron transfer to the sugar-phosphate backbone, leading to bond cleavage via DEA.[45] Due to the structural complexity of the biological environment around RNA and associated histone proteins, simplified model systems comprising an isolated nucleobase and an amino acid can be used to study resonance behavior. While such models effectively capture local electronic interactions, they overlook critical aspects of the full biological context, including solvent effects, secondary structure, and macromolecular interactions. However, too many details in complex models make it impossible to perform practical calculations and often obscure the big picture. Therefore, we employed a simple model to obtain insight into the effect of amino acids on nucleobase-centered resonances in RNA. Glycine is one of the smallest amino acids available. This study aims to understand the effect of binding the amino acid glycine to uracil-centered anionic resonances.

## 2. Computation details:

We have used the Resonance via Padé (RVP) method[73–75,88] put forth by the Moiseyev group to obtain the resonance positions and widths from the stabilization graph. The DLPNO-based implementation of the EA-EOM-CCSD (EA-EOM-DLPNO-CCSD)[89,90] method with the TIGHTPNO setting has been used for the electronic structure calculations. The resonance calculations for the nucleobase and the microsolvated complexes are performed using the cc-pVTZ[91] basis set augmented with 2s, 2p, and 2d diffuse functions on heavy atoms. The notation cc-pVTZ+2s,2p,2d denotes the expanded basis set. The additional functions are generated in an even-tempered way using a scaling factor of 2.0. Subsequently, their exponents were scaled by dividing with the parameter ($\alpha$), such that $\alpha>1$ yields a more diffuse basis and $\alpha<1$ results in a more compact basis. The relevant auxiliary basis set was selected using the Autoaux tool of ORCA software. All the calculations to construct stabilization plots are performed using ORCA 5.0.3.[92,93] The comprehensive stabilization graphs for isolated uracil and glycine, plotted at the EA-EOM-DLPNO-CCSD/RVP method, are provided in Figure 1, showcasing dependence on the scaling parameter α. The resonance wave function behaves like a localized state near the interaction region and is largely insensitive to small perturbations from basis set scaling. In the stabilization plot, this results in minimal energy variation with respect to the scaling parameter α, in contrast to continuum states, which exhibit significant energy shifts with changes in α. The calculations using the RVP method are performed using an open-source "Automatic RVP" software.[73,94] More details of the protocol for resonance calculations using RVP and EA-EOM-DLPNO-CCSD method can be found in the reference.[52]

The geometries of uracil and glycine in the gas phase are optimized at the RI-MP2[95,96]/def2-TZVP[97] level of theory. The low-energy conformers of the microsolvated uracil complexes are generated via conformational sampling with the help of CREST software.[98] The energetically favorable conformer for each complex is selected and further optimized at the RI-MP2/def2-TZVP level of theory for subsequent calculations. The optimized geometries for uracil and the microsolvated uracil complexes considered in this study are provided in Figure S1, and the corresponding Cartesian coordinates are provided in the supporting information.

## 3. Results and Discussion:

### 3.1 Shape resonances in isolated Uracil and Glycine

We have obtained three low-lying shape resonances for our model pyrimidine nucleobase uracil anion. The resulting resonances are systematically characterized and are designated as the $1\pi^*$, $2\pi^*$, and $3\pi^*$ states, corresponding to increasing energies. The stabilization plot for the uracil anion with highlighted $\pi^*$ resonances, is provided in Figure 1A. Table 1 presents the calculated resonance positions and associated widths for the three shape-type resonance states ($1\pi^*$, $2\pi^*$, and $3\pi^*$) of the uracil anion, as obtained using the EA-EOM-DLPNO-CCSD/RVP method. The first two $\pi^*$ shape resonances of uracil anion appear at energies of approximately 0.55 eV and 2.03 eV, with associated lifetimes of around 94 fs and 21 fs, respectively. These states exhibit relatively narrow widths, particularly the first resonance, which has a width of only ~0.007 eV, indicating a long-lived quasi-bound nature. The third resonance appears at 4.78 eV and is significantly broader and short-lived with a lifetime of approximately 1.8 fs. Previous studies on uracil anion shape resonances have also consistently overestimated the energy of the third $\pi^*$ resonance, which is considerably higher than the experimentally reported value by Aflatooni et al. as 3.83 eV[56] and by Cooper et al. as 3.89±0.03 eV.[99,100] This discrepancy highlights a persistent difficulty in the precise characterization of high-lying resonances.

To evaluate the accuracy and reliability of our computational approach, these results are compared with previously reported theoretical predictions and available experimental data (See Table 1). The resonance positions of all three uracil anion shape resonances in the present study are in good agreement with the results obtained using the RVP-EA-EOM-CCSD method by Bouskila et al.[101]. Our results show a consistent underestimation of resonance position by approximately 0.1–0.2 eV across all the resonance states when compared with the values derived from the analytically continued EA-EOM-CCSD method using the Generalized Padé Approximation (GPA) approach by Fennimore and Matsika.[28] Furthermore, our calculated resonance energies are consistent with those obtained through the SAC-CI method with complex absorbing potentials (CAP), as reported by Sommerfield and Ehara.[64] The resonance positions obtained from our RVP-EA-EOM-DLPNO-CCSD calculations are higher than experimental values, but they agree well with previously published theoretical studies. This suggests that the present approach is reliable for estimating the energies of shape resonances in nucleobases. However, our results tend to underestimate the widths of these resonances. The previously reported values for the width of uracil anion shape resonances show considerable deviation among them, the only general agreement being that the width tends to increase with the resonance energy. Unfortunately, there is no experimental data available for the widths of these resonance states. It should be noted that the calculated resonance position and width depend upon three factors[102]: the underlying electronic structure method, the $L^2$ methods used to calculate the complex energy, and the basis set used for the calculations. The variation of one or more of these factors can lead to significant differences in width among different studies.

Before analyzing the influence of glycine on uracil shape resonances, it is important to investigate the resonance states of the isolated glycine anion. Electron attachment to glycine can result in the formation of metastable anionic states, which have been characterized both experimentally and theoretically. Experimental studies by Aflatooni et al.[103] using electron transmission spectroscopy (ETS) reported a low-lying π* shape resonance in glycine, primarily localized on the –COOH group, with a resonance energy of 1.93 eV. Gianturco and co-workers'[104] quantum scattering calculations identified four resonances for glycine, one with A″ symmetry at 3.32 eV and three with the A′ symmetry at 8.85, 8.95, and 14.38 eV. Varella and co-workers[105] employed the Schwinger multichannel (SMC) method and reported a π* shape resonance in the 2.3−2.8 eV energy range and an A′ shape resonance around 9.5 eV. Tashiro's R-matrix calculations[106] detected a significant peak in the A′ partial cross-section at roughly 3.4 eV energy, originating from a temporary trapping of the scattering electron in the π* orbital of the −COOH group. In this study, we have identified a π*−type (2.51 eV) and a σ*−type (9.81 eV) shape resonances of glycine anion from the stabilization plot at EA-EOM-DLPNO-CCSD/RVP level (See Table 2). The stabilization plot and the natural orbitals associated with the native glycine are provided in Figure 1 (B). The π* and σ* shape resonance states have lifetimes of 23 fs and 2 fs, respectively. The low-lying π* shape resonance of glycine is of particular interest due to its energy and lifetime being comparable to the 2π* shape resonance of uracil. The electron density in this state is largely localized on the carboxyl group of the amino acid. One could expect this state to interact with the 2π* resonance of uracil. This may lead to delocalization of the excess electron across both moieties and potentially modify the electronic structure and lifetime of the uracil-centered resonance. This interaction is important for understanding the influence of a protein-like environment on nucleobase-centered anionic resonances.

## 3.2 Effect of microsolvation with glycine (uracil(Gly))

During the folding and unfolding processes of RNA, amino acid residues from adjacent proteins or peptides can interact closely with nucleobases through non-covalent interactions, including hydrogen bonding, electrostatic forces, and π–π interactions. While basic residues like lysine and arginine often bind to the phosphate backbone, polar amino acids can interact directly with the nucleobases. To probe the influence of amino acid environments on nucleobase-centered anionic resonances, we have investigated the shape resonance states of uracil anion in complex with glycine (uracil(Gly)) as a representative model system. This model system allows us to systematically examine how interactions with amino acid residues affect the position and lifetime of π* resonance states in nucleobases.

Our calculations show that the presence of glycine significantly stabilizes the π* shape resonance states of the uracil anion nucleobase. In the uracil(Gly) anionic complex, the 1π*, 2π*, and 3π* resonances are shifted to lower energies compared to their positions in isolated uracil anion (see Table 3 and Figure S2). The 2π* state exhibits electron density primarily localized on the uracil nucleobase, with minor delocalization onto the glycine moiety (Figure 2). The 3π* state shows the most pronounced redshift (~0.30 eV), likely arising from limitations of the basis set in capturing high-lying resonance states. A distinct π*(glycine) resonance appears at 2.21 eV, absent in the isolated uracil anion. The natural orbital analysis of this π*(glycine) resonance state (Figure 2) shows that the excess electron density is predominantly localized on the glycine molecule, with a small density distribution over the uracil nucleobase. This spatial distribution of electron density suggests that the presence of glycine significantly influences the resonance behavior of uracil. From a molecular orbital (MO) perspective, the formation of the additional resonance state observed in the uracil–glycine anionic complex can be attributed to the interaction between the π* molecular orbital of the glycine anion and the low-lying 2π* orbital of the uracil anion. The interaction of molecular orbitals results in the formation of two distinct states: one predominantly localized on uracil and the other primarily localized on glycine. This indicates a coupling between the π* systems of uracil and glycine. Therefore, the interaction between the orbitals leads to energy lowering of the uracil-centered resonance states, effectively stabilizing them relative to their counterparts in isolated uracil anion. We are unable to locate the glycine σ* resonance in the stabilization plot for uracil(Gly). This may be due to the shift of the σ* resonance state of glycine to high energy values. The lifetime of 1π* state has been increased by factor of 3.5 upon complexation. The 2π* and π*(glycine) states differ by ~0.5 eV in energy and ~3 fs in lifetime. Whereas the

3π* state exhibits a lifetime of 3 fs, about 1.2 fs longer than in the uracil anion. These findings demonstrate the significant influence of glycine on the resonance properties of the uracil anion.

### 3.3 Effect of microhydration (uracil(H$_2$O))

Water molecules constitute a significant part of the cellular environment and play a crucial role in modulating the behavior of biomolecules. Therefore, to gain a more comprehensive understanding of environmental effects on nucleobase-centered anionic resonances, it is important to compare the stabilization effects induced by microsolvation with amino acids to those arising from simple hydration. In this context, we have compared the stabilization of anionic π* resonances in the uracil–glycine anionic complex with that in monohydrated uracil anion. The microsolvation of uracil anion by a single water molecule leads to the stabilization of all three π* shape resonance states (See Table 3 and Figure S2). However, the qualitative nature of the three resonance states remained unaffected by the presence of a water molecule, as shown by the corresponding natural orbitals depicted in Figure 2. The stabilization effects of microhydration on the shape resonances of the uracil anion are clearly reflected in the redshifts observed in their resonance energies. The 1π* and 2π* states exhibit energy shifts of approximately 0.04 eV and 0.05 eV, respectively, while the 3π* state shows a redshift of around 0.26 eV compared to isolated uracil anion. The 3π* state also exhibits slight electron density distribution on the water molecule. In addition to the redshift in energy, an increase in the lifetimes is observed for all three resonances for the monohydrated uracil complex compared to the uracil anion. Natural orbital analysis confirms that the electron density in all three resonance states remains primarily localized on the uracil nucleobase. These results suggest that a single water molecule does not significantly alter the qualitative nature of the lowest three shape resonances of the uracil anion.

### 3.4 Effect of microsolvation with zwitterionic glycine (uracil(zGly))

We anticipate that the protective effect demonstrated by amino acids in the thin films will also extend to biological media. However, in the biological environment, amino acids such as glycine may adopt a zwitterionic form. Therefore, it is essential to understand the effect of glycine zwitterion on the TNIs of uracil. Before discussing the shape resonances of the uracil-glycine zwitterion (uracil(zGly)) anion complex, it is important to investigate the difference between the anionic resonance states of glycine in its native and zwitterionic forms. A comparative analysis of the two states can aid in elucidating the role of intramolecular charge separation in modifying the electronic structure and stabilizing the resonance states. Since the zwitterionic form is not stable in the gas phase, we have used its di-hydrated geometry (zGly.2H$_2$O) in our calculations. The stabilization plot for zGly.2H$_2$O is shown in Figure S3, where the π* and σ* states are identified and highlighted. The π* resonance state of zGly.2H$_2$O (Figure S4 B) is blue-shifted by approximately 1.0 eV, while the σ* resonance is red-shifted and has an energy of 6.43 eV, which is around 3.4 eV lower than that of the native glycine anion (See Table S1). The lifetime of both states in the zwitterionic form is also reduced when compared to the neutral form.

The optimized geometry of the anionic uracil-zwitterionic glycine complex (uracil(zGly)) is shown in Figure S1. Clearly, the nature of the hydrogen bonding interaction between glycine and uracil in uracil(zGly) is different from uracil(Gly). This is because of the absence of the carboxylic proton in uracil(zGly). Uracil-zwitterionic glycine anionic complex has four π* shape resonances, similar to the uracil(Gly) complex. The natural orbitals of the resonance states are represented in Figure 3, which also includes the glycine-centered π* state (π*(zglycine)) (Figure S4 A). The nucleobase-centered 2π* resonance in uracil(zGly) shows a very similar resonance position as that observed in uracil(Gly) (Table S1). A significant distinction between the complexes of native and zwitterionic glycine is observed in the 3π* resonance. In the uracil(zGly) complex, the electron density is delocalized across both uracil and glycine, while in the uracil(Gly) complex, it is solely localized on uracil. Nevertheless, the presence of glycine, irrespective of its native or zwitterionic forms, introduces a new resonance state where the electron density is delocalized over the glycine. This can reduce the possibility of the incoming electron from getting predominantly localized over uracil upon electron attachment. The trend observed from uracil(Gly) and uracil(zGly) suggests that glycine-centered states may assume a larger dominant role as glycine concentration increases.

## 4. Conclusions

In this work, we have simulated the effect of the amino acid environment on the microsolvated RNA-pyrimidine base model system using accurate quantum chemistry methods. Our work establishes that the effect of the glycine environment on the uracil shape resonance states is more prominent than that observed in aqueous media. Glycine not only stabilizes the uracil-centered resonance in the TNI but also adds its new resonance states where the electron density distribution is primarily localized over glycine itself. These results point out that in an amino acid-rich environment, the amino acids can capture the electron density and move it away from the nucleobase through the formation of an amino acid-centered TNI. The present study on the uracil-glycine anion radical model systems shows a significant interaction between the nucleobase-centered resonances and the amino acid. Thus, these results could provide us with possible clues on the role of amino acids in the processes involved in the radiation damage to RNA. Of course, we note that the model considered in the present initial study is rather simplistic and may not accurately describe the behavior of RNA in the physiological environment. In addition to the nucleobase shape resonances, the core-excited resonances could also play a major role in DEA. Therefore, one should also analyze the influence of high electron energy ($\geq 3$ eV) on a protein-rich environment of nucleobases. Work is in progress towards this direction.

**Supplementary Material**

Cartesian coordinates of the microsolvated structures, microsolvation stabilization graphs, and optimized geometries are provided in the supplementary material.

**Acknowledgments**

The authors acknowledge the support from the IIT Bombay, CRG (Project no. CRG/2018/001549), and Matrix project of DST-SERB (Project No. MTR/2021/000420), CSIR-India (Project No. 01(3035)/21/EMR-II), UGC-India, DST-Inspire Faculty Fellowship (Project no. DST/INSPIRE/04/2017/001730), Prime Minister's Research Fellowship, ISRO (Project No. RD/0122-ISROC00-004) for financial support. IIT Bombay super computational facility, and C-DAC Supercomputing resources (Param Smriti, Param Brahma) for computational time. AA acknowledges the support from the NSF under Grant No. CHE-1920110. A.A. also thanks Université Paris-Saclay for the Visiting Professorship at the Institut de Chimie Physique and support from the Free Radical Research and Radiation Biology Program, the University of Iowa. AKD acknowledges the research fellowship funded by the EU NextGenerationEU through the Recovery and Resilience Plan for Slovakia under project No. 09I03-03-V04-00117.

**Table and Figures:**

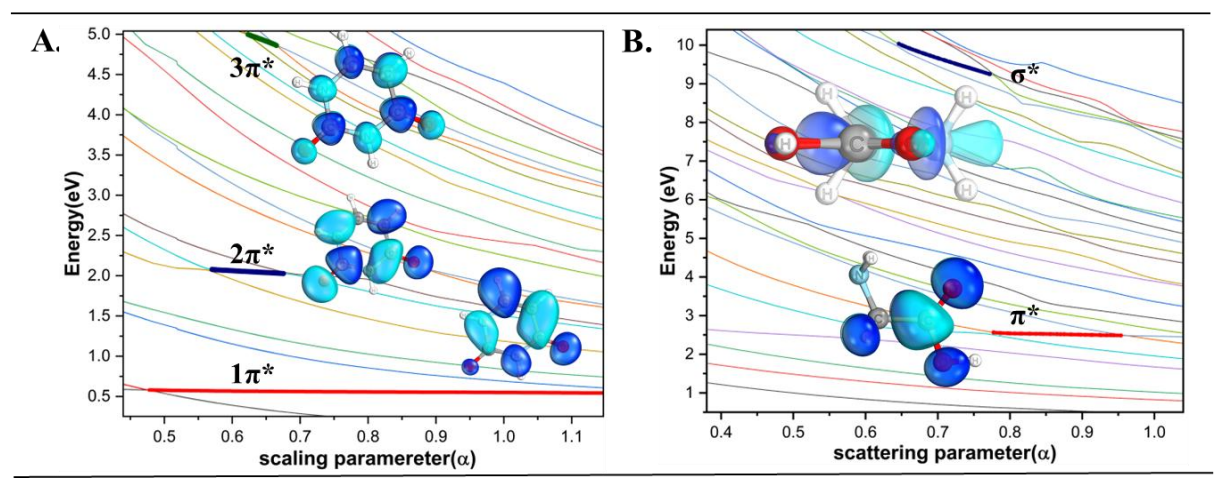

**Figure 1:** *Stabilization plots of A. isolated uracil anion and B. native glycine anion with the highlighted stable regions and natural orbitals corresponding to the shape resonances.*

**Table 1**: *The comparison of resonance position energies and widths (in parentheses) for uracil with previous experimental and theoretical results. Values are given in eV.*

| Resonance state | 1π* | 2π* | 3π* |
|---|---|---|---|
| Method | | | |
| **RVP-EA-EOM-DLPNO-CCSD**[a] | 0.55 (0.007) | 2.03 (0.032) | 4.78 (0.375) |
| **RVP-EA-EOM-CCSD**[101] | 0.597 (0.014) | 2.183 (0.14) | 4.858 (0.657) |
| **GPA-EA-EOM-CCSD**[28] | 0.61 (0.02) | 2.28 (0.07) | 4.98 (0.34) |
| **CAP-SAC-CI**[64] | 0.57 (0.05) | 2.21 (0.10) | 4.82 (0.58) |
| **S-KT,S-KB/DFT**[107] | 0.36 (0.05) | 1.75 (0.10) | 4.52 (0.23) |
| **R matrix**[108] | 0.13 (0.003) | 1.94 (0.17) | 4.95 (0.38) |
| **R matrix**[109] | 2.27 (0.21) | 3.51 (0.38) | 6.50 (1.03) |
| **SMCPP**[110] | 0.14 (0.005) | 1.76 (0.15) | 4.83 (0.78) |
| **Expt.**[56] | 0.22 | 1.58 | 3.83 |

[a] This work.

**Table 2:** *Resonance position (eV) and width (eV) corresponding to the shape resonance states of native glycine anion calculated using the RVP-EA-EOM-DLPNO-CCSD/cc-pVTZ+2s2p2d level of theory*

| Glycine | $E_R$ | $\Gamma$ |
|---|---|---|
| $\pi^*$ | 2.51 | 0.029 |
| $\sigma^*$ | 9.81 | 0.370 |
| $\pi^*$ (Exp.)[103] | 1.93 | |

**Table 3:** *The effect of microsolvation on the resonance position (eV) and width (eV) of shape resonance states of uracil anion calculated using RVP-EA-EOM-DLPNO-CCSD/cc-pVTZ+2s2p2d level of theory.*

| Molecule | cc-pVTZ+2s2p2d | |
|---|---|---|
| Uracil | $E_R$ | $\Gamma$ |
| $1\pi^*$ | 0.55 | 0.007 |
| $2\pi^*$ | 2.03 | 0.032 |
| $3\pi^*$ | 4.78 | 0.375 |
| Uracil($H_2O$) | | |
| $1\pi^*$ | 0.51 | 0.004 |
| $2\pi^*$ | 1.98 | 0.029 |
| $3\pi^*$ | 4.52 | 0.329 |
| Uracil(Gly) | | |
| $1\pi^*$ | 0.50 | 0.002 |
| $2\pi^*$ | 1.68 | 0.030 |
| $\pi^*$(glycine) | 2.21 | 0.036 |
| $3\pi^*$ | 4.48 | 0.230 |

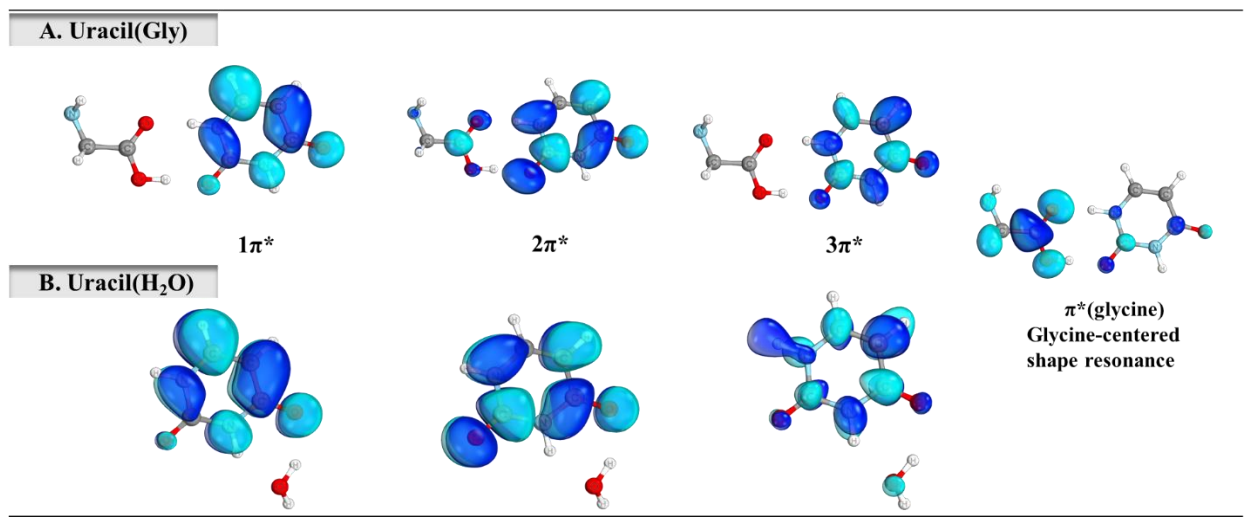

**Figure 2:** *Natural orbitals corresponding to π\* shape resonances of A) uracil-glycine (uracil(Gly)) and B) uracil-water (uracil(H$_2$O) anionic complexes. (Right: Glycine-centered resonance from uracil(Gly))*

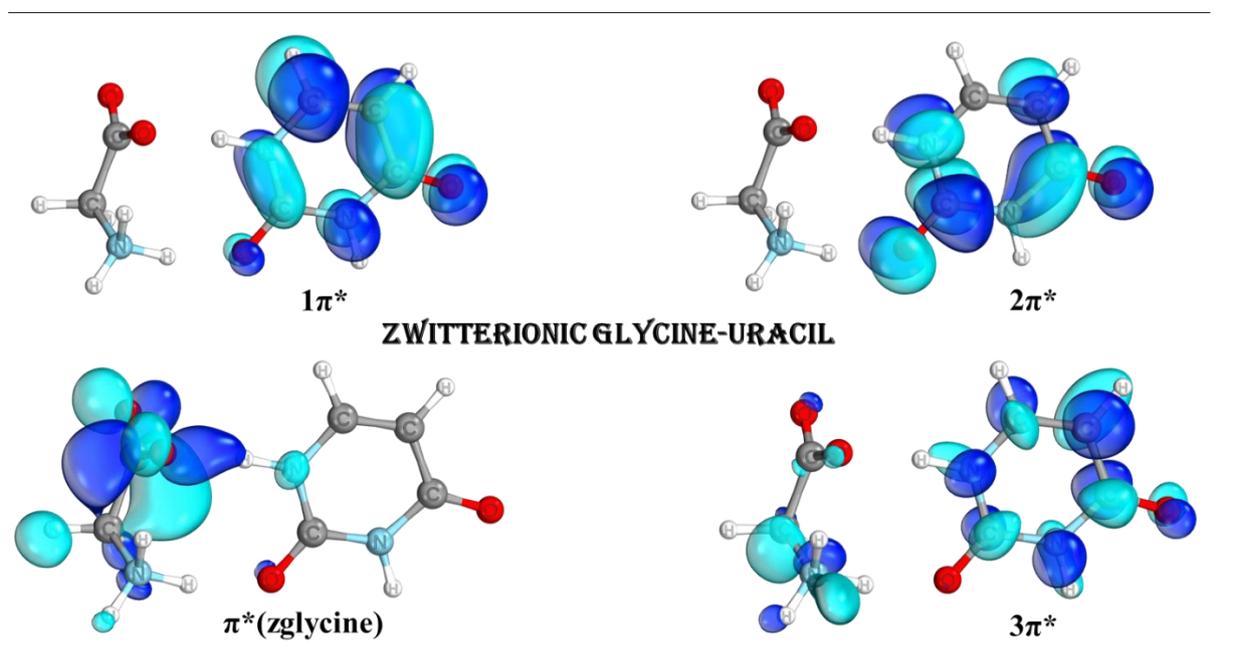

**Figure 3:** *Natural orbitals corresponding to the zwitterionic glycine-uracil anionic complex (uracil(zGly)).*